\def\kms{\ifmmode{\,\hbox{km}\,s^{-1}}\else {\rm\,km\,s$^{-1}$}\fi}

\def\msun{{\rm\,M_\odot}}
\def\lsun{{\rm\,L_\odot}}

\def\kmsm{{\rm\,km\,s^{-1}\,Mpc^{-1}}}

\def\hmpc{\ifmmode{h^{-1}\,\hbox{Mpc}}\else{$h^{-1}$\thinspace Mpc}\fi}

\def\eg{{\it e.g.}~}

\def\et{{\it et~al.}~}

\def\sigp{\ifmmode{\sigma_p}\else {$\sigma_p$}\fi}
\def\sig1{\ifmmode{\sigma_1}\else {$\sigma_1$}\fi}
\def\r200{\ifmmode{r_{200}}\else {$r_{200}$}\fi}
\def\rp200{\ifmmode{r^\prime_{200}}\else {$r^\prime_{200}$}\fi}

\def\mlobs{289\pm50}

\def\mlclose{1136\pm138}
\def\omzeroc{0.19\pm0.06}
\def\afit{0.66\pm0.09}
\def\mvadj{0.82\pm0.14}
\def\ladj{0.11\pm0.05} %mags
\def\sigall{1.05\pm0.04}
\def\rvall{3.35\pm0.32}
\def\afitblu{1.82\pm0.27}
\def\afitred{0.56\pm0.10}

\def\sigblu{1.27\pm0.11}
\def\sigred{0.97\pm0.04}
\def\rvblu{5.28\pm0.73}
\def\rvred{2.58\pm0.27}
\def\mvblu{2.31\pm0.42} 
\def\mvred{0.66\pm0.08} %ratio is 3.5
\def\ratiosig{1.31\pm0.13}
\def\ratiorv{2.05\pm0.34}
\def\ratiomv{3.5\pm1.3}

\documentstyle[11pt,aaspp4]{article}
\tighten

\begin{document}

\title
{The Dynamical Equilibrium of Galaxy Clusters}

\author{
R.~G.~Carlberg\altaffilmark{1,2},
H.~K.~C.~Yee\altaffilmark{1,2},
E.~Ellingson\altaffilmark{1,3},
S.~L.~Morris\altaffilmark{1,4},
R.~Abraham\altaffilmark{1,4,5},
P.~Gravel\altaffilmark{1,2},
C.~J.~Pritchet\altaffilmark{1,6},
T.~Smecker-Hane\altaffilmark{1,4,7}
F.~D.~A.~Hartwick\altaffilmark{6}
J.~E.~Hesser\altaffilmark{4},
J.~B.~Hutchings\altaffilmark{4},
\& J.~B.~Oke\altaffilmark{4}
}

\altaffiltext{1}{Visiting Astronomer, Canada--France--Hawaii Telescope, 
	which is operated by the National Research Council of Canada,
	le Centre National de Recherche Scientifique, and the University of
	Hawaii.}
\altaffiltext{2}{Department of Astronomy, University of Toronto, 
	Toronto ON, M5S~3H8 Canada}
\altaffiltext{3}{Center for Astrophysics \& Space Astronomy,
	University of Colorado, CO 80309, USA}
\altaffiltext{4}{Dominion Astrophysical Observatory, 
	Herzberg Institute of Astrophysics,	
	National Research Council of Canada,
	5071 West Saanich Road,
	Victoria, BC, V8X~4M6, Canada}
\altaffiltext{5}{Institute of Astronomy, 
	Madingley Road, Cambridge CB3~OHA, UK}
\altaffiltext{6}{Department of Physics \& Astronomy,
	University of Victoria,
	Victoria, BC, V8W~3P6, Canada}
\altaffiltext{7}{Department of Physics \& Astronomy,
	University of California, Irvine,
	CA 92717, USA}

%\clearpage

\begin{abstract}
If a galaxy cluster is effectively in dynamical equilibrium then all
galaxy populations within the cluster must have distributions in
velocity and position that individually reflect the same underlying
mass distribution, although the derived virial masses can be quite
different. Specifically, within the CNOC cluster sample the virial
radius of the red galaxy population is, on the average, a factor of
$\ratiorv$ smaller than that of the blue population. The red galaxies
also have a smaller RMS velocity dispersion, a factor of $\ratiosig$
within our sample.  Consequently, the virial mass calculated from the
blue galaxies is $\ratiomv$ times larger than from the red galaxies.
However, applying the Jeans equation of stellar-hydrodynamical
equilibrium to the red and blue subsamples separately give
statistically identical cluster mass profiles.  This is strong
evidence that these clusters are effectively equilibrium systems, and
therefore empirically demonstrates that the masses in the virialized
region are reliably estimated using dynamical techniques.
\end{abstract}

\keywords{galaxies: clusters, cosmology: large-scale structure of universe}

%\clearpage

\section{Introduction}

The primary goal of the CNOC (Canadian Network for Observational
Cosmology) cluster redshift survey is to obtain a value of density
parameter, $\Omega_0$, for those components of the mass field
(possibly all) that participate in gravitational clustering (Carlberg
\et\ 1996, hereafter CNOCi). The product of the field luminosity
density, $j$, with its mass-to-light ratio, $M/L$, estimates the mean
mass density of the universe, $\rho_0$ (\cite{oort}), which in ratio
to the critical density, $\rho_c$ is equal to $\Omega_0$ (\cite{gt}),
when calculated in current epoch co-ordinates.  The crucial quantity
is an estimator of the field mass-to-light ratio, for which we use the
cluster virial mass-to-light ratio, $M_v/L$. The main technical
concern of the CNOC survey is that there are various possibilities
that $M_v$ is a biased estimator of the cluster mass.  Besides
statistical biases, dissipative processes can cause the cluster
galaxies to become more centrally concentrated than the cluster
mass. In addition, when estimating the field $M/L$ from the cluster
results, allowance must be made for the differences between the
cluster and field galaxy populations.

To address these issues a sample of high X-ray luminosity galaxy
clusters were selected from the EMSS catalogue (\cite{emss,gl}).  This
produces a relatively homogenous set of 16 clusters that likely
contain a substantial virialized component. The sample, at a mean
redshift of about \onethird, gives a substantial column in redshift
space which we use to calculate the field luminosity density in
precisely the same measurement system as the cluster light.  The
galaxies were selected with no knowledge of the colour or whether they
were cluster members, although great care is taken to control the
selection effects that are unavoidable to maintain efficiency
(\cite{yec}).

From these data we calculated the line-of-sight velocity dispersions,
$\sigma_1$, and virial radii, $r_v$, of the clusters to derive the
virial masses (\cite{bt_gd}), $GM_v=3\sigma_1^2 r_v$. The velocity
dispersions are calculated with an iterative technique that
statistically removes interlopers by taking advantage of the
relatively large foreground and background sample to estimate
interloper densities (\cite{global}). This technique leads to velocity
dispersions that are on the average 7\% less than those found from
precisely the same data in the redshift range, and about 13\% less
with a somewhat less restrictive redshift cut, even when evaluated
with the iterated bi-weight estimator (\cite{bfg}).  The virial radius
is calculated from the data in the redshift range of the cluster with
a ``ring-wise'' potential estimator, which allows correction for the
strip sampling and helps reduce the noise (details are given in
\cite{global}).  All these global quantities are derived from the
individual galaxy positions, velocities, colors and luminosities, for
which objective errors are calculated using the Jackknife method
(\cite{et}).  The errors are sufficiently small that the
assumption of a symmetric distribution is appropriate and is verified
by comparing to Bootstrap error estimates.

The average cluster virial mass-to-light ($r$ band, k-corrected) ratio
is $\langle{M_v/L_r^k}\rangle=\mlobs h
\msun/\lsun$ (\cite{global}). In the field, over the same redshift 
range, with the same sampling and corrections, the closure
mass-to-light ratio is $\rho_c/j=\mlclose h\msun/\lsun$ (in co-moving
units, for $q_0=0.1$).  There are two additional corrections applied
to $\langle{M_v/L_r^k}\rangle$.  The mean luminosity of galaxies
brighter than $M_r^k$ of $-18.5$ mag is $\ladj$ lower in the cluster
than the field and the analysis of the mass profile (Carlberg, Yee \&
Ellingson 1997, hereafter \cite{profile}) showed that $M_v$ needed to
be multiplied by $\mvadj$, which we attribute to neglecting the
surface term in the virial mass estimator.  Making these corrections
leads to a redshift adjusted $\Omega_0=\omzeroc$.

Our conclusion that the stellar-hydrodynamical (SHD) mass profile of
the clusters averaged together closely follows the number density
profile of the galaxies (\cite{profile}) rests on the assumption that
the clusters are effectively in equilibrium such that the Jeans
equation can be used to derive the mass profile from the radial and
velocity distribution of any tracer population. In this paper we
demonstrate that two radically different tracer populations are
consistent with the same underlying mass profile, which we take as
strong support of the equilibrium assumption.  In the following
section we describe the division of the sample into two completely
disjoint blue and red subsamples, whose surface density and projected
velocity dispersion profiles are measured and demonstrate the gross
differences in the calculated virial masses. In Section 3 we derive
the stellar-hydrodynamical potential generating mass profiles in which
the galaxies are orbiting. The implications of the statistical
equality of the two profiles are discussed in the final section. All
calculations in this paper assume $H_0=100\kmsm$ and $q_0=0.1$,
although the results are either independent or not sensitive to these
choices.

\section{The Blue and Red Subsamples}

The CNOC cluster redshift survey catalogues (\cite{yec}) contain
$\sim$2600 galaxies having Gunn $r$ magnitudes, $g-r$ colors, and
redshifts. The sample is split at the color $(g-r)_z = (g-r)
[1.2+2.33(z-0.3)]^{-1}$ of $0.7$ (\cite{global}) . This is
approximately the color of an Sab galaxy, which roughly divides
galaxies into spheroid dominated, ``old'' stellar systems and disk
dominated, ``young'' stellar systems.  Fourteen clusters are averaged
together after removing two ``binary'' clusters.

To a limit of $M_r^k=-18.5$ mag, about 70\% of the cluster galaxies
fall into the red subsample.  The methods of analysis are precisely
the same as used previously for the full sample (\cite{profile}). The
velocities are normalized to the $\sigma_1$ of each cluster calculated
from galaxies of all colors so that the two subsamples have a common
reference.  The RMS velocity dispersions of the red and blue galaxies
are $\sigred$ and $\sigblu$, respectively, in units where the
velocities are normalized to the velocity dispersions of all the
galaxies in each cluster. That is, the velocity dispersion of blue
cluster galaxies is about 30\% higher than for red galaxies
(\cite{rpkk,kg,stein}). The projected radii are normalized to
$\r200=\sqrt{3}\sigma_1/(10H(z))$, the radius at which the mean
interior overdensity in a cluster is $200\rho_c(z)$. The quantity
$H(z)$ is the Hubble constant at the redshift of the cluster.  The
virial radii, $r_v$, are $\rvred$ and $\rvblu$ for red and blue
galaxies, respectively, in \r200\ units.  In ratio to the sample as a
whole, the average virial masses are $\mvred$ and $\mvblu$,
respectively, for red and blue subsamples. The ratio of these two
virial masses is $\ratiomv$. As is well known, the red galaxies in
clusters are much more centrally concentrated than the blue galaxies
(\eg\ \cite{dressler}).  Our data usefully illustrate that the
galaxy color selection can make a very large difference to the
estimated mass. In general, blue or emission line selected galaxies
will always give higher mass estimates than absorption line galaxies.

The projected surface number density profiles, $\Sigma_N(R)$ of the
two subsamples are displayed in Figure~\ref{fig:surf}.  The profiles
are fitted to the projection of the density function
(\cite{hernquist}),
\begin{equation}
\nu(r) = {A\over{r(r+a)^3}},
\label{eq:nu}
\end{equation}
by adjusting $A$ and $a$ to minimize the residuals.  The fits give
scale radii $a$ of $\afitred$ and $\afitblu$ for red and blue
populations, respectively. For the sample as a whole $a=\afit$. In
both cases the $\chi^2$ per degree of freedom is about 0.8, with the
errors estimated as the average of the upper and lower $1\sigma$
confidence range from the bootstrap analysis.

The observed projected velocity dispersion, $\sigp(R)$, can
be used to infer the real space radial velocity dispersion,
$\sigma_r(r)$, via,
\begin{equation}
\sigp^2(R)\Sigma_N(R) = \int_R^\infty \nu(r)\sigma_r^2
        (1-\beta{R^2\over r^2}) {r\over\sqrt{r^2-R^2}} \,dr,
\label{eq:sigp}
\end{equation}
where $\beta=1-\sigma_\theta^2/\sigma_r^2$ is the velocity anisotropy
parameter. The parameters of 
\begin{equation}
\sigma_r^2(r)= {B\over{b+r}},
\label{eq:sigfit}
\end{equation}
are adjusted until the $\chi^2$ with the observed $\sigp(R)$ are
minimized.  The data and the fits for two values of $\beta$ are
displayed in Figure~\ref{fig:sig}. Again the errors are from the
symmetrized bootstrap confidence range.

\section{The Mass Distribution}

The mass distribution in which the galaxies are in orbital equilibrium
can be inferred with no assumption about the relative distribution of
mass and galaxies from the projected velocity dispersion profile,
$\sigp(R)$, and the projected galaxy density distribution,
$\Sigma_N(R)$. Remarkably, we found (\cite{profile}) that the
integrated galaxy number density profile, $L(r)$, of our full sample
is statistically indistinguishable from the integrated mass profile,
$M(r)$.   

Mass profiles are derived from the fitted $\nu(r)$ and $\sigma_r(r)$
using the Jeans equation (\cite{bt_gd}),
\begin{equation}
M_{SHD}(r) = -{\sigma_r^2r\over G}
        \left[{{d \ln{\sigma_r^2}}\over{d\ln{r}}} +
        {{d\ln{\nu}}\over{{d\ln{r}}}} +2\beta\right].
\label{eq:jeans}
\end{equation}
We will call this mass the stellar hydrodynamical mass. 
The resulting $M_{SHD}(r)$ from the red and blue galaxies are
shown in Figure~\ref{fig:mol} for $\beta=0$ and $\onehalf$.
The quantity plotted is the virial mass-to-light bias,
\begin{equation}
b_{Mv}(r) = {{M_{SHD}(r)}\over{L(r)}}  {\tilde{L} \over\tilde{M}_v},
\label{eq:mlbias}
\end{equation}
where $\tilde{L}$ is an arbitrary normalization of the total light
that cancels between numerator and denominator. The quantity
$\tilde{M}_v$ is the normalization of the virial mass of the
dimensionless sample as a whole,
$\tilde{M}_v=3\tilde{\sig1}^2\tilde{r}_v$, where
$\tilde{\sig1}=\sigall$ and $\tilde{r}_v=\rvall$. If the virial mass
calculated from the sample as a whole gave the correct $M/L$ at all
radii then $b_{Mv}(r)=1$ everywhere. If the tracer population follows
the true mass profile but with some scale error, then $b_{Mv}$ will be
a constant other than unity.  Figure~\ref{fig:mol} shows that beyond
about $ 0.3 \r200$ the mass profiles deduced from the two subsamples
are identical within their errors (about 15 and 30\% for red and blue
mass profiles, respectively).  Figure~\ref{fig:mvbias} shows the value
of $b_{Mv}$ evaluated at \r200, demonstrating that the blue and red
subsamples give statistically identical values. Furthermore the value
is always less than unity, as we found for the sample as a whole
(\cite{profile}) which has the implication that although the galaxy
numbers fairly accurately trace the cluster mass profile the virial
mass to light ratio is always an overestimate of the mean $M/L$ inside
\r200.  There is no statistically significant radial gradient of
$b_{Mv}(r)$ for $r\simeq\r200$.

There are several important conclusions to be drawn from
Figure~\ref{fig:mol}.  First, it provides strong evidence that the
blue and red galaxies are sufficiently in equilibrium with the cluster
mass distribution that these dynamical methods work to recover the
true mass profile. Second, the fitted scale radius of the mass
profile (taken to be equal to that of the full sample)
$a=\afit$, is slightly more extended than the red galaxies (\cite{tf})
$a=\afitred$, but is substantially more compact than the blue galaxy
distribution, $a=\afitblu$, all measured in
\r200\ units. The differences in scale radii imply that neither the
blue nor the red galaxies are distributed like the mass, so the virial
masses calculated from these subsamples are not correct.

\section{Conclusions}

The densities and velocity dispersions of the red and blue galaxy
subsamples, although very different from each other, imply a
statistically identical mass profile for the cluster potential. This
is necessary if both populations are in equilibrium with the
potential, validating our use of the Jeans equation.  This mass
profile is statistically identical to the number density distribution
of all the galaxies of our full sample.  The limiting factor in the
precise numerical agreement is that there are relatively few blue
galaxies in clusters, meaning that their density and velocity
dispersion profiles are less accurately measured.

There are problems and issues that cannot be addressed with a sample of
this size. A much larger sample would likely find that a common value
of $\beta$ would not work for both red and blue galaxies, since it is
expected that on the average blue galaxies will avoid the core, hence
near the centre have smaller $\beta$ than the red galaxies, and at
large radius include many objects on their first cluster crossing,
hence have higher $\beta$ than the red galaxies. Furthermore the
galaxies themselves are not invariant mass points. The blue galaxies
are being altered by the cluster environment such that some of the
their members are likely leaving the blue population to join the red
population (\cite{a2390}).

The virial mass calculated from the
full sample, empirically adjusted for its measurement biases, will
correctly estimate the mass enclosed. With these two subsamples we
have now tested each step in the chain of logic which supports our
corrected, population adjusted value of $\Omega_0=\omzeroc$. There is
no compelling evidence for any remaining systematic errors of the
cluster $M/L$ as an estimator of the field value over the redshift
range of this sample. 

\acknowledgments
We thank the Canadian Time Assignment Committee of the CFHT for
allocations of observing time, and the CFHT organization for the
technical support which made these observations feasible.  Funding was
provided by NSERC and NRC of Canada.

\newpage

\figcaption[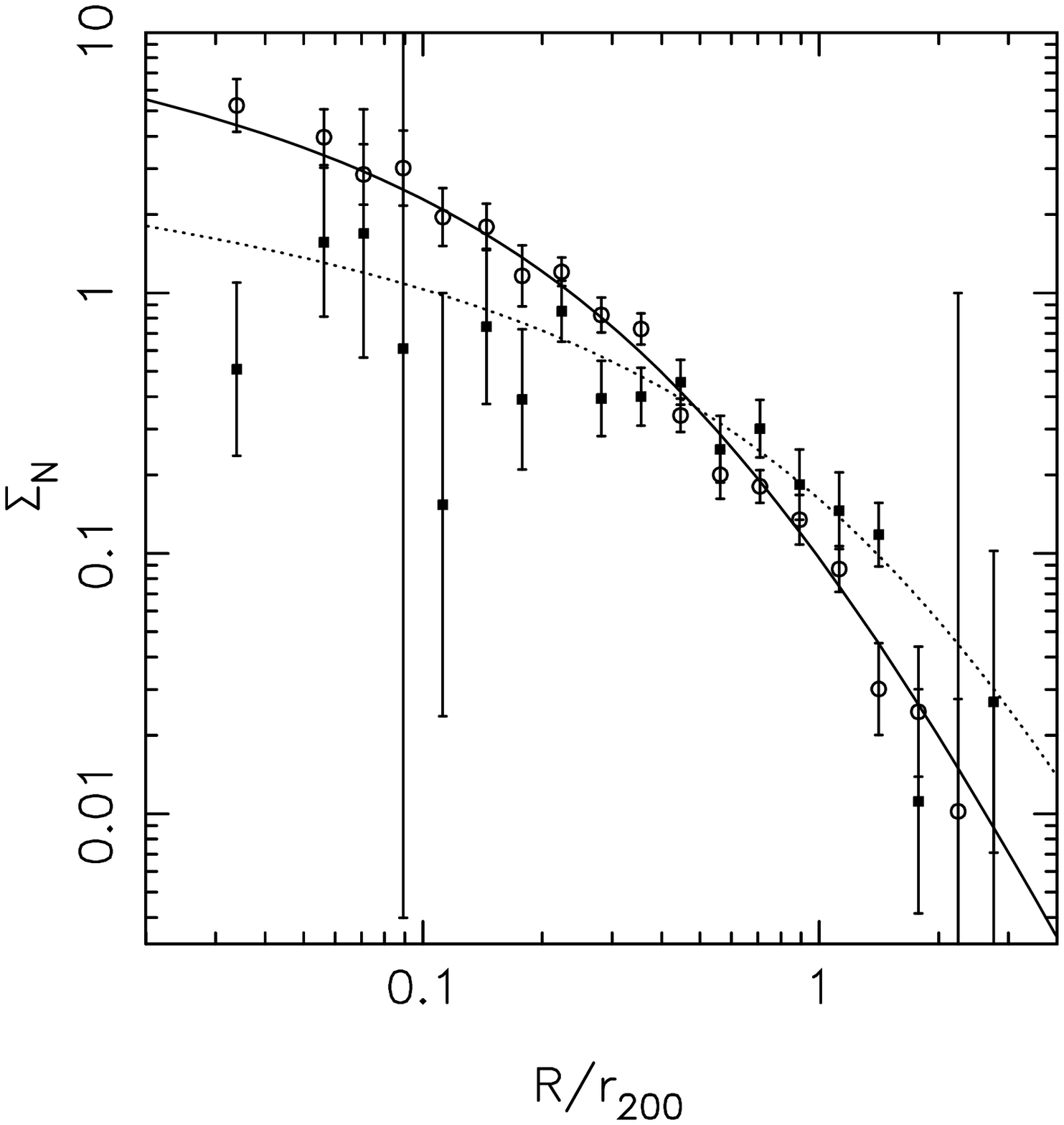]{The projected number density profiles of
the blue (solid squares and dotted lines) and red (circles and solid lines)
cluster galaxies.  The $1\sigma$ confidence range from a Bootstrap
error estimate is shown. \label{fig:surf}}

\figcaption[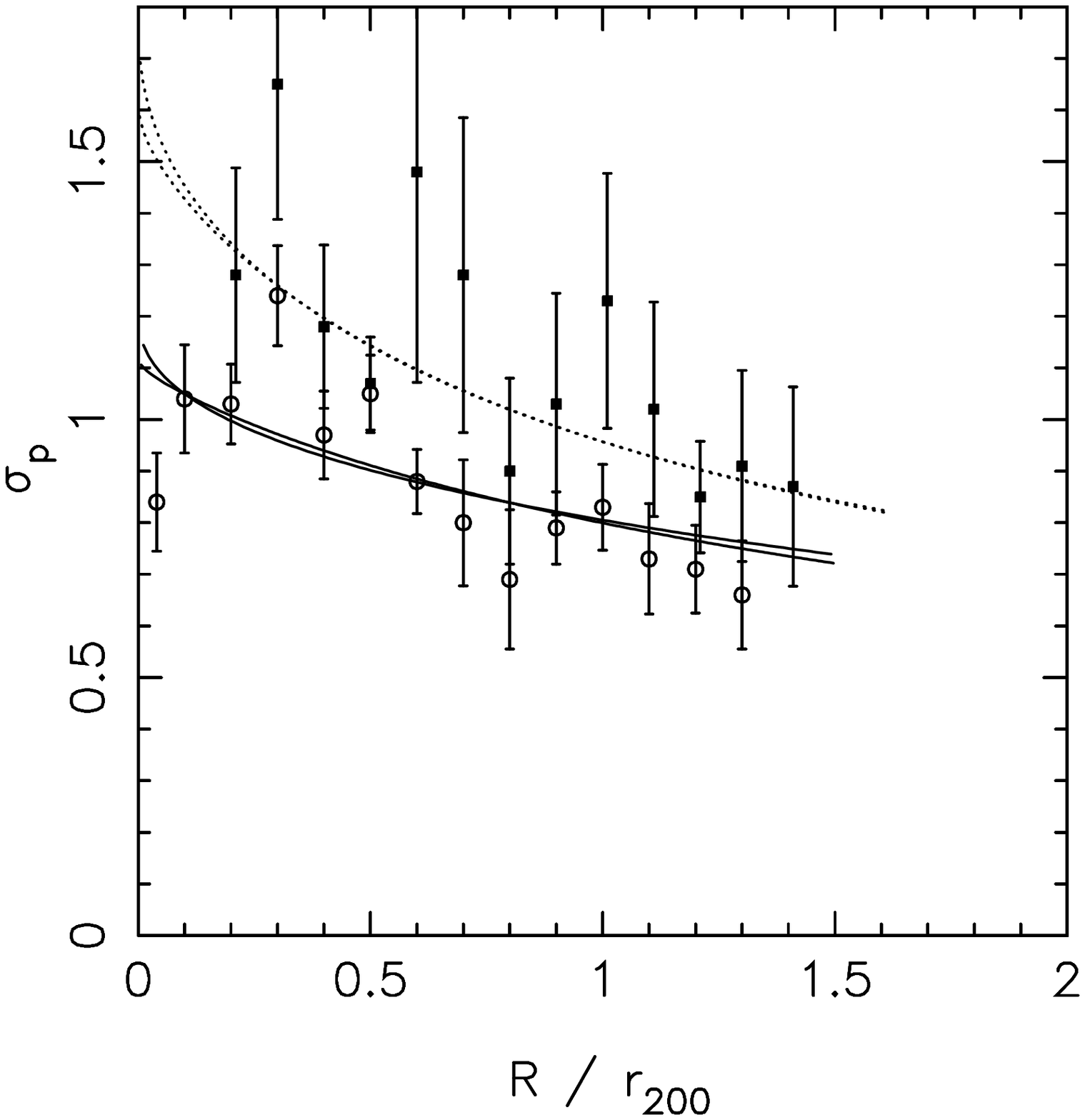]{The RMS line-of-sight velocity dispersion, 
\sigp, as a function of projected radius.
The blue galaxies (solid squares and dotted line) have a larger \sigp\ than
the red ones (circles and solid line). The $\sigp(R)$ are fitted with
$\beta=[0.0,0.5]$.
\label{fig:sig}}

\figcaption[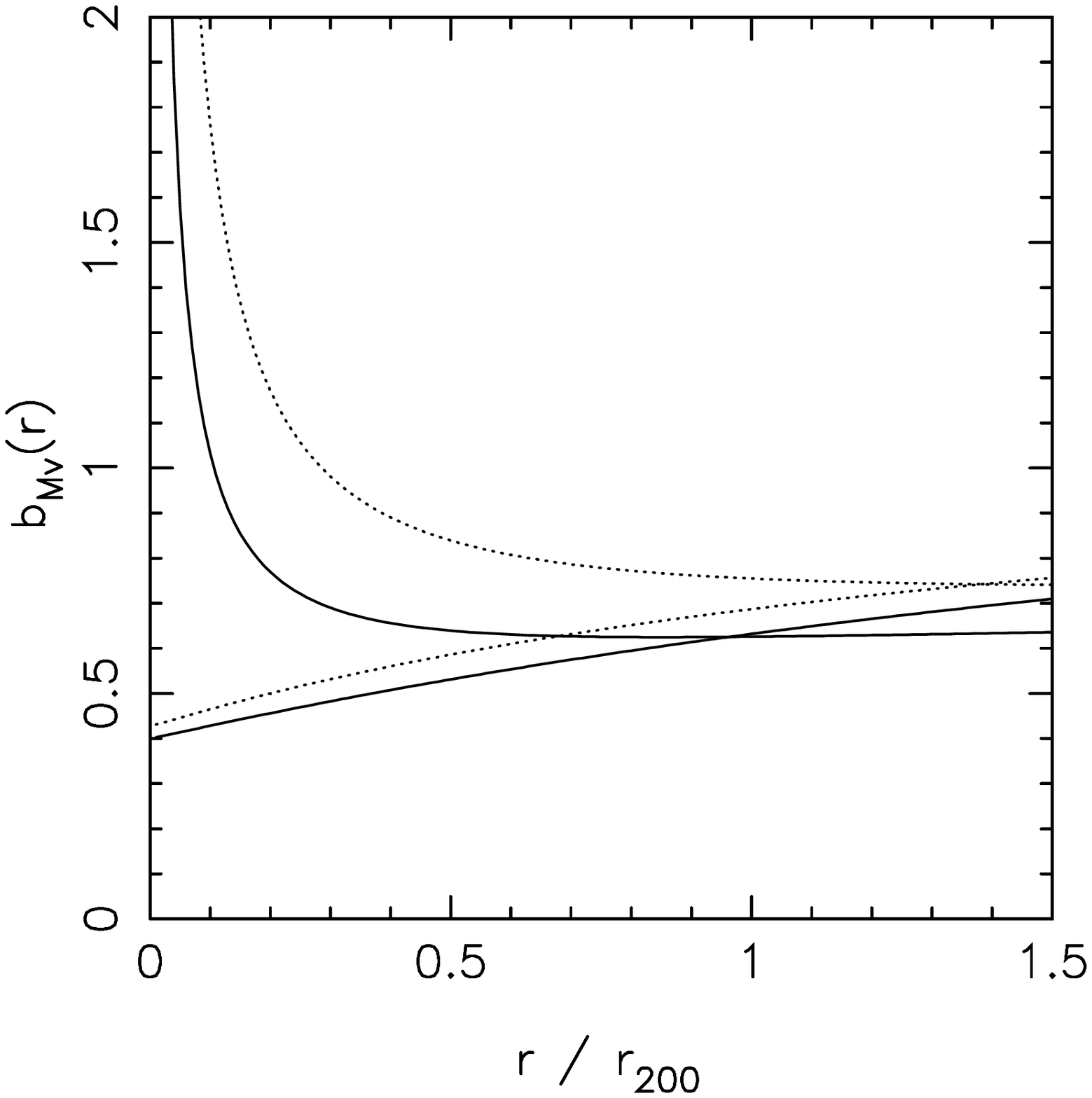]{The bias function refered to
the {\em total} light profile, $b_{Mv}(r)$ calculated from the mass
profiles derived from blue (dotted line) and red (solid line)
subsamples.  The upper line for both subsamples is for $\beta=0$, the
lower line is for $\beta=1/2$.
\label{fig:mol}}

\figcaption[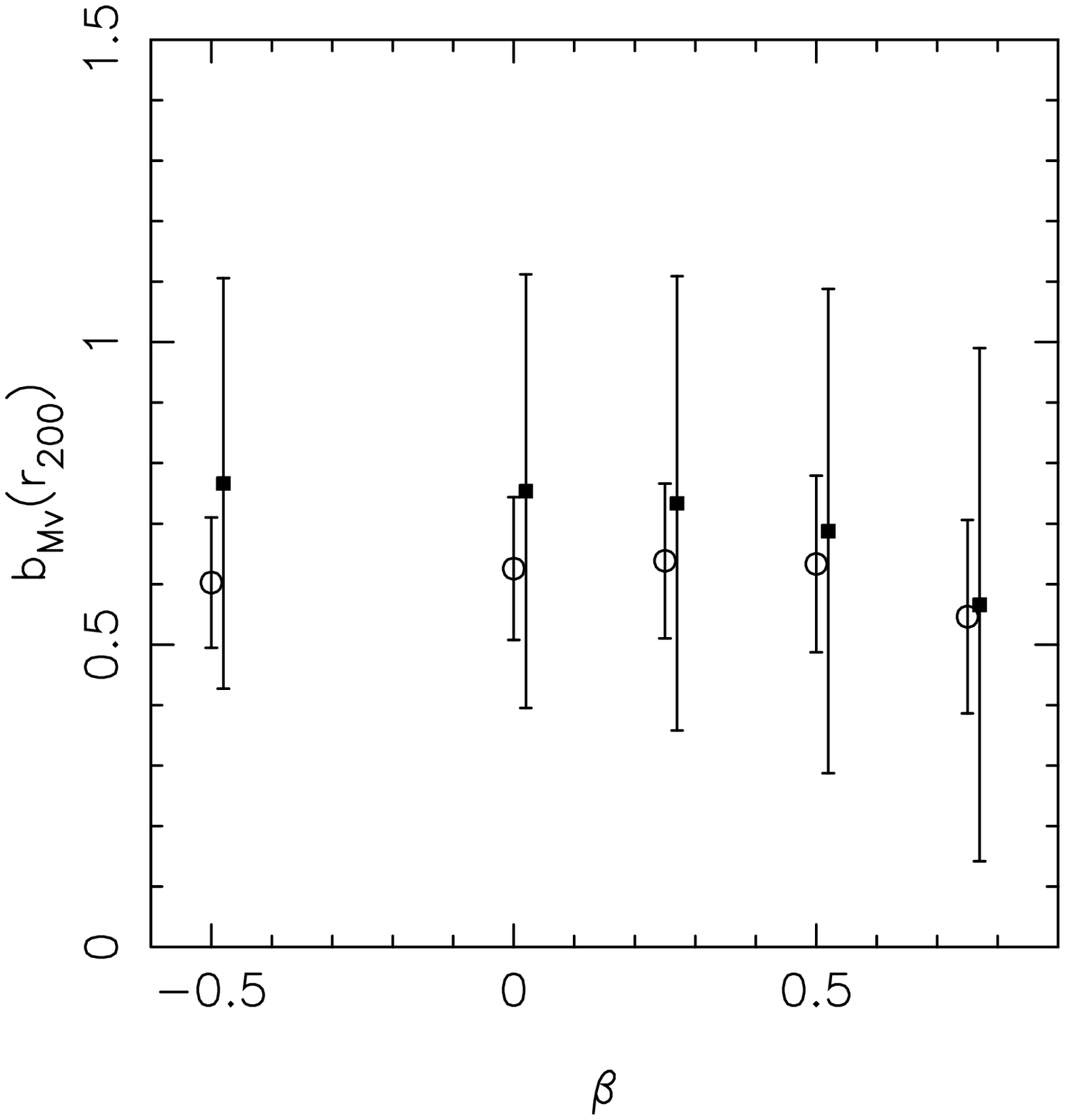]{The value of $b_{Mv}$ evaluated at
\r200. The blue subsample results are denoted with solid squares
and the red with open circles.\label{fig:mvbias}}

\begin{figure}[h]\epsscale{0.9}\figurenum{1}\plotone{fig1.ps}\caption{}
	\end{figure}\epsscale{1.0}
\begin{figure}[h]\figurenum{2}\plotone{fig2.ps}\caption{}\end{figure}
\begin{figure}[h]\figurenum{3}\plotone{fig3.ps}\caption{}\end{figure}
\begin{figure}[h]\figurenum{4}\plotone{fig4.ps}\caption{}\end{figure}

\end{document}